\documentclass[a4paper,11pt]{article}%
\usepackage{geometry}
\usepackage{amsmath}
\usepackage{amsfonts}
\usepackage{amssymb}
\usepackage{graphicx}
\usepackage{indentfirst}
\usepackage[small,bf]{caption}
\usepackage{slashed}
\usepackage[utf8]{inputenc}
\usepackage{caption}
\usepackage{subcaption}
\providecommand{\U}[1]{\protect\rule{.1in}{.1in}}

\begin{document}

\date{}
\title{\textbf{ Gribov horizon and non-perturbative BRST symmetry in the maximal Abelian gauge}}
\author{\textbf{M.~A.~L.~Capri}\thanks{caprimarcio@gmail.com}\,\,,
\textbf{D.~Fiorentini }\thanks{diegofiorentinia@gmail.com}\,\,,
\textbf{S.~P.~Sorella}\thanks{silvio.sorella@gmail.com}\,\,\,,\\[2mm]
{\small \textnormal{  \it Departamento de F\'{\i }sica Te\'{o}rica, Instituto de F\'{\i }sica, UERJ - Universidade do Estado do Rio de Janeiro,}}
 \\ \small \textnormal{ \it Rua S\~{a}o Francisco Xavier 524, 20550-013 Maracan\~{a}, Rio de Janeiro, Brasil}\normalsize}
	 
\maketitle

\begin{abstract}
The non-perturbative nilpotent exact BRST symmetry of the Gribov-Zwanziger action in the Landau gauge constructed in \cite{Capri:2015ixa} is generalized to the case of Euclidean  Yang-Mills theories quantized in the maximal Abelian gauge. The resulting diagonal gluon propagator is evaluating in dimensions $D=4,3,2$. In $D=4,3$ a decoupling type behaviour is found in the infrared region, while in $D=2$ a scaling type behaviour emerges.

\end{abstract}

\section{Introduction}

Since the seminal work \cite{Gribov:1977wm}, the issue of the Gribov copies has become a powerful tool in order to investigate the behaviour of confining Yang-Mills theories in the 
non-perturbative infrared region. The deep progress \cite{Gribov:1977wm,Singer:1978dk,Zwanziger:1988jt,Zwanziger:1989mf,Zwanziger:1992qr,Dell'Antonio:1991xt,vanBaal:1991zw}  done in the last four decades on the Gribov problem\footnote{For a pedagogical introduction to the Gribov problem, see \cite{Sobreiro:2005ec,Vandersickel:2012tz}.  }  in the Landau gauge, has resulted in the so-called Gribov-Zwanziger framework \cite{Gribov:1977wm,Zwanziger:1988jt,Zwanziger:1989mf,Zwanziger:1992qr} and its refined version \cite{Dudal:2007cw,Dudal:2008sp,Dudal:2011gd}, from which a local and renormalizable non-perturbative action taking into account the existence of Gribov copies has been constructed. So far, this set up has provided a large number of applications which have covered several aspects of confining Yang-Mills theories, namely: study of the gluon and ghost correlation functions \cite{Zwanziger:1988jt,Zwanziger:1989mf,Zwanziger:1992qr,Dudal:2007cw,Dudal:2008sp,Dudal:2011gd,Cucchieri:2012cb}, investigation of the spectrum of the glueballs \cite{Dudal:2010cd,Dudal:2013wja},  thermodynamics and phase transitions \cite{Canfora:2015yia,Canfora:2013kma,Canfora:2013zna,Lichtenegger:2008mh,Fukushima:2012qa,Fukushima:2013xsa,Su:2014rma}, supersymmetric theories  \cite{Capri:2014tta,Capri:2014xea},  study of the confinement/deconfinement transition when Higgs fields are present \cite{Capri:2012ah}. \\\\Parallel to these developments, an intense and rich discussion on the important aspect of the relationship between the BRST symmetry and the Gribov problem has taken place, see \cite{Dudal:2009xh,Sorella:2009vt,Baulieu:2008fy,Capri:2010hb,Dudal:2012sb,Dudal:2014rxa,Capri:2012wx,Pereira:2013aza,Pereira:2014apa,Capri:2014bsa,Tissier:2010ts,Serreau:2012cg,Serreau:2015yna,Lavrov:2013boa,Moshin:2015gsa,Schaden:2014bea,Schaden:2015uua,Cucchieri:2014via} for an overview. Needless to say, the BRST symmetry is a fundamental ingredient of the Faddeev-Popov quantization, allowing for a perturbative  all order proof of the renormalizability of Yang-Mills theories as well as for the identification of the asymptotic Fock  sub-spaces on which the $S$-matrix\footnote{We refer here to non-confining Yang-Mills theories for which the elementary fields admit asymptotic states, so that a  perturbative construction of the $S$-matrix can be worked out.} is unitarity. Unravelling the role of the BRST symmetry in the case of  confining theories is believed to be a pivotal issue in order to understand the mechanism responsible for color confinement. \\\\Recently, the authors \cite{Capri:2015ixa} have been able to construct a non-perturbative nilpotent extension of the standard BRST operator which turns out to be an exact symmetry of the Gribov-Zwanziger action in Landau gauge.  This is an important step towards a comprehension of the role of this symmetry at the non-perturbative level. The construction of the non-perturbative BRST operator outlined in \cite{Capri:2015ixa} has allowed for a nice and geometrical resolution of the Gribov problem in the class of the linear covariant gauges, with the important outcome that the correlation functions of the gauge invariant colorless operators are independent of the gauge parameter, a feature also shared by the so-called Gribov parameter $\gamma^2$. These are non-trivial results, given the well known difficulties of addressing the Gribov issue in the class of the linear covariant gauges  \cite{Sobreiro:2005vn,Capri:2015pja}, due to the lack of hermiticity of the Faddeev-Popov operator. \\\\In this work, we extend the construction of \cite{Capri:2015ixa} to the case of the maximal Abelian gauge, which plays a central role in the dual superconductivity mechanism for confinement \cite{'tHooft:1981ht,Nambu:1974zg,Mandelstam:1974pi}. According to the dual superconductivity picture, $QCD$ at low energies should behave as an Abelian theory in presence of monopoles, a feature  referred as to Abelian dominance 
\cite{Ezawa:bf,Sasaki:1998ww,Sasaki:1998th,Suzuki:1995va,Suzuki:1989gp,Shiba:1994ab,Stack:1994wm,Hioki:1991ai,Miyamura:1995xn,Amemiya:1998jz,Bornyakov:2003ee,Gongyo:2013sha,Sakumichi:2014xpa,Mendes:2007vv,Mendes:2008vf}. The condensation of the monopoles would give rise to the formation of flux tubes which  confine quarks \cite{'tHooft:1981ht,Nambu:1974zg,Mandelstam:1974pi}. \\\\Concerning the Gribov problem in the maximal Abelian gauge, although the situation cannot  be compared to that of the Landau gauge \cite{Gribov:1977wm,Zwanziger:1988jt,Zwanziger:1989mf,Zwanziger:1992qr,Dell'Antonio:1991xt}, a few results are already available, see \cite{Capri:2005tj,Capri:2006cz,Capri:2007hw,Capri:2008ak,Capri:2008vk,Capri:2010an,Capri:2011ki}, where the analogous of Zwanziger's horizon function  as well as of the  Gribov-Zwanziger action and of its refined version have been constructed. A study of the maximal Abelian gauge within the context of the Schwinger-Dyson equations can be found in \cite{Huber:2009wh}. \\\\The paper is organized as follows. In Sect.2 we give a short summary of the construction of the non-perturbative BRST symmetry in the case of the Gribov-Zwanziger action in the Landau gauge. Sect.3 is devoted to the generalization to the maximal Abelian gauge. In Sect.4 we discuss the refined action and its non-perturbative BRST symmetry. In Sect.5 we analyse the resulting diagonal gluon propagator in $D=4,3,2$ dimensions pointing out that, similarly to what happens in the case of the Landau gauge, while a decoupling behaviour takes place in $D=4,3$, a scaling type behaviour emerges in $D=2$. In Sect.6 we provide a detailed account of how the non-perturbative BRST symmetry and associated action can be cast in local form by introducing a suitable set of localizing fields. In Sect.7 we collect our conclusion.

\section{Brief summary of the Gribov-Zwanziger framework and of its non-perturbative BRST symmetry in Landau gauge} 

For the benefit of the reader, let us give here a short summary of the Gribov-Zwanziger set up for $SU(N)$ Euclidean Yang-Mills theory in the Landau gauge,  $\partial_\mu A^a_\mu=0$, and of its non-perturbative BRST symmetry.  This framework \cite{Gribov:1977wm,Zwanziger:1988jt,Zwanziger:1989mf,Zwanziger:1992qr,Sobreiro:2005ec,Vandersickel:2012tz} enables us to implement the restriction in the functional integral to the Gribov region $\Omega$, defined as 
\begin{equation}
\Omega = \{ \; A^a_\mu|  \;  \partial_\mu A^a_\mu =0,  \;    {\mathcal M}^{ab}(A) > 0 \; \}    \;, \label{om}
\end{equation}
where ${\mathcal{M}}^{ab}$ is the Faddeev-Popov operator of the Landau gauge, {\it i.e.}
\begin{equation}
\mathcal{M}^{ab}(A)=-\delta^{ab}\partial^2+gf^{abc}A^{c}_{\mu}\partial_{\mu},\,\,\,\, \mathrm{with}\,\,\,\, \partial_{\mu}A^{a}_{\mu}=0\,.
\label{intro0}
\end{equation}
The restriction to the region $\Omega$ takes into account the existence of the Gribov copies which affect the Faddeev-Popov quantization scheme \cite{Gribov:1977wm,Zwanziger:1988jt,Zwanziger:1989mf,Zwanziger:1992qr,Sobreiro:2005ec,Vandersickel:2012tz}. 
According to \cite{Gribov:1977wm,Zwanziger:1988jt,Zwanziger:1989mf,Zwanziger:1992qr},  for the partition function of quantized Yang-Mills action in Landau gauge,  one writes 
\begin{equation}
{Z}=\int_{\Omega} \left[{D}{A}\right]\; \delta(\partial A) \; \det({\cal M})\;\mathrm{e}^{- S_{\mathrm{YM}} }\,.
\label{intro1}
\end{equation}
The restriction of the domain of integration to the region $\Omega$ can be effectively implemented by adding to the starting action an additional non-local term $H(A)$, known as the horizon function. More precisely \cite{Gribov:1977wm,Zwanziger:1988jt,Zwanziger:1989mf,Zwanziger:1992qr}
\begin{eqnarray}
&&\int_{\Omega} \left[DA\right]\; \delta(\partial A) \; \det({\cal M})\;\mathrm{e}^{- S_{\mathrm{YM}} }\nonumber\\&&  = \int \left[{D}{A}\right]\; \delta(\partial A) \; \det({\cal M})\;\mathrm{e}^{- \left( S_{\mathrm{YM}} +\gamma^4H(A) - 4V\gamma^4(N^2-1) \right) }
\label{gz1}
\end{eqnarray}
where
\begin{equation}
H(A)=g^2\int d^4xd^4y~f^{abc}A^{b}_{\mu}(x)\left[\mathcal{M}^{-1}(x,y)\right]^{ad}f^{dec}A^{e}_{\mu}(y)\,,
\label{intro3}
\end{equation}
with $\left[\mathcal{M}^{-1}\right]$ denoting the inverse of the Faddeev-Popov operator, eq.\eqref{intro0}.
The mass parameter  $\gamma^2$ appearing in expression \eqref{gz1}  is known as the Gribov parameter. It  is determined in a self-consistent way by the  gap equation \cite{Gribov:1977wm,Zwanziger:1988jt,Zwanziger:1989mf,Zwanziger:1992qr}

\begin{equation}
\langle H \rangle = 4V(N^2-1)\,,
\label{intro4}
\end{equation}
where the vacuum expectation values $\langle H \rangle$ has to be evaluated with the measure defined in eq.\eqref{gz1}; $V$ denotes the space-time volume. Expression \eqref{gz1} can be cast in a more suitable form by introducing a set of commuting $({\bar \phi}, \phi)$ and anticommuting $(\omega, {\bar \omega})$ auxiliary fields \cite{Zwanziger:1988jt,Zwanziger:1989mf,Zwanziger:1992qr}, namely
\begin{equation}
\int_{\Omega} \left[{D}{A}\right]\; \delta(\partial A) \; \det({\cal M})\;\mathrm{e}^{- S_{\mathrm{YM}} } = \int  \left[{D}{{\Phi}}\right]\;\mathrm{e}^{- \left( S_{GZ}  - 4V\gamma^4(N^2-1) \right)},
\label{gzact1}
\end{equation}
where $\Phi$ refers to all fields present and $S_{GZ}$ stands for the Gribov-Zwanziger action, {\it i.e.}
\begin{equation}
S_{GZ} = S_{FP} + \int d^4x\;  \left( {\bar \phi}^{ac}_\mu  [{{\cal M}}(A)]^{ab} \phi^{bc}_\mu  - {\bar \omega}^{ac}_\mu [ {{\cal M}}(A)]^{ab}  \omega^{bc}_\mu + g \gamma^2 f^{abc}A^a_\mu ({\bar \phi}^{bc}_\mu + \phi^{bc}_\mu)   \right)\;, \label{gzact}
\end{equation}
with $S_{FP}$ denoting the Faddeev-Popov action of the Landau gauge
\begin{equation}
S_{FP} = S_{YM} + \int d^4x \left( b^a \partial_\mu A^a_\mu + {\bar c}^a \partial_\mu D^{ab}_\mu c^b \right)    \;.   \label{fp}
\end{equation}
Notice that, in the expression \eqref{fp} above, the field $b^{a}$ denotes  the Lagrange multiplier enforcing the Landau gauge, while $c^{a}$ and $\bar{c}^{a}$ are the Faddeev-Popov ghost and anti-ghost fields, respectively.\\\\In the local formulation, the gap equation \eqref{intro4} can be rewritten as
\begin{equation}
\frac{\partial{\mathcal E}_v}{\partial \gamma^2}=0 \;, \qquad  e^{-V{\mathcal E}_v} = \int  \left[{D}{{\Phi}}\right]\;\mathrm{e}^{- \left( S_{GZ}  - 4V\gamma^4(N^2-1) \right)}  \;, \label{gapr}
\end{equation}
where ${\mathcal E}_v$ denotes the vacuum energy.  \\\\In order to construct a non-perturbative BRST symmetry for the action  \eqref{gzact}, we follow  \cite{Capri:2015ixa} and introduce the non-local transverse gauge invariant field\footnote{We employ here a matrix notation, $A^h_\mu= A^{h,a}_\mu T^a$, where $\{ T^a \}$, $a=1,...(N^2-1)$,  denote the hermitian generators of $SU(N)$, $[T^a,T^b]=if^{abc}T^c$.} $A_{\mu}^h$, $\partial_\mu A^h_\mu=0$,  \cite{Zwanziger:1990tn, Lavelle:1995ty,Capri:2005dy} 
\begin{eqnarray}
A_{\mu }^{h} &=&P_{\mu\nu} \left( A_{\nu }-ig\left[ \frac{\partial A}{\partial
^{2}},A_{\nu }\right] +\frac{ig}{2}\left[ \frac{\partial A}{\partial ^{2}}%
,\partial _{\nu }\frac{\partial A}{\partial ^{2}}\right] \right)
+O(A^{3})  \nonumber \\
&=&A_{\mu }-\frac{\partial _{\mu }}{\partial ^{2}}\partial A+ig\left[ A_{\mu
},\frac{1}{\partial ^{2}}\partial A\right] +\frac{ig}{2}\left[ \frac{1}{%
\partial ^{2}}\partial A,\partial _{\mu }\frac{1}{\partial ^{2}}\partial
A\right]\nonumber\\&& +ig\frac{\partial _{\mu }}{\partial ^{2}}\left[ \frac{\partial
_{\nu }}{\partial ^{2}}\partial A,A_{\nu }\right]   +i\frac{g}{2}\frac{\partial _{\mu }}{\partial ^{2}}\left[ \frac{\partial A%
}{\partial ^{2}},\partial A\right] +O(A^{3})\,,  \label{hhh3g}
\end{eqnarray}
where $P_{\mu\nu}=\left(\delta _{\mu \nu }-\frac{\partial _{\mu }\partial
_{\nu }}{\partial ^{2}}\right) $ is the transverse projector. Expression \eqref{hhh3g} is left invariant by infinitesimal gauge transformations order by order, namely 
\begin{equation}  
\delta A^h_\mu =0  \;, \qquad \delta A_\mu = -\partial \lambda +ig [A_\mu, \lambda] \;. \label{ginvah} 
\end{equation}
Looking now at eq.\eqref{hhh3g}, one sees that a divergence $\partial A$ is present in all higher order terms. Therefore,  we can rewrite Zwanziger's horizon function $H(A)$ in terms of the invariant field $A^h$ as
\begin{equation}
H(A) = H(A^h) - R(A) (\partial A)
\end{equation}
where $R(A)(\partial A)$ is a short-hand notation, $R(A) (\partial A)= \int d^4x d^4y R^a(x,y) (\partial A^a)_y$, $R(A)$ being an infinite non-local power series of $A_\mu$. Thus, for the Gribov-Zwanziger action, we may write
\begin{eqnarray}
S_{GZ} &=& S_{YM} + \int d^4x \left( b^a \partial_\mu A^a_\mu + {\bar c}^a \partial_\mu D^{ab}_\mu c^b \right)  + \gamma^4 H(A)     \nonumber \\
&&\hspace{-1cm} = ~S_{YM} + \int d^4x \left( b^a \partial_\mu A^a_\mu + {\bar c}^a \partial_\mu D^{ab}_\mu c^b  \right)  + \gamma^4 H(A^h) -\gamma^4 R(A) (\partial A)     \nonumber \\
& = & S_{YM} + \int d^4x \left( b^{ah} \partial_\mu A^a_\mu + {\bar c}^a \partial_\mu D^{ab}_\mu c^c \right)  + \gamma^4 H(A^h)      \;,
\label{gzh2}
\end{eqnarray}
where the new field $b^{ah}$ stands for
\begin{equation}
b^h = b - \gamma^4 R(A)   \;. \label{bh}
\end{equation}
The use of the field $b^h$ enables us to write down an exact nilpotent non-perturbative BRST symmetry. Rewriting the Gribov-Zwanziger action by using the auxiliary  fields $({\bar \phi}, \phi, \omega, {\bar \omega})$, {\it i.e.}
\begin{eqnarray}
S_{GZ} &=& S_{YM} + \int d^4x \left( b^{ah} \partial_\mu A^a_\mu + {\bar c}^a \partial_\mu D^{ab}_\mu c^b \right) \nonumber\\&& \hspace{-0.9cm}+ \int d^4x\;  \left( {\bar \phi}^{ac}_\mu [{{\cal M}}(A^h)]^{ab} \phi^{bc}_\mu  - {\bar \omega}^{ac}_\mu [{{\cal M}}(A^h)]^{ab} \omega^{bc}_\mu + g\gamma^2 f^{abc} A^{ah}_\mu ({\bar \phi}^{bc}_\mu + \phi^{bc}_\mu)   \right), \label{gzh3}
\end{eqnarray}
it turns out that  expression \eqref{gzh3} is left invariant by the nilpotent non-perturbative BRST transformation  \cite{Capri:2015ixa}
\begin{eqnarray}
s_{\gamma^2} A^a_\mu &=& - D_\mu^{ab} c^b \;, \qquad s_{\gamma^2}  c^a = \frac{g}{2} f^{abc} c^b c^c\;,  \nonumber \\ 
s_{\gamma^2} {\bar c}^a  &=&   b^{ah}   \;, \qquad  s_{\gamma^2} b^{ah} = 0      \;, \nonumber \\
s_{\gamma^2}  \phi_\mu^{ab} & = & \omega_\mu^{ab} \;, \qquad  s_{\gamma^2}  \omega_\mu^{ab}=0 \;, \nonumber \\
s_{\gamma^2} {\bar \omega}_\mu^{ab} &=&   {\bar \phi}_\mu^{ab}  +  \gamma^2 gf^{kpb} A_\mu^{h,k}\left[{\cal M}^{-1}(A^h)\right]^{pa}  \;, \qquad   s_{\gamma^2} {\bar \phi}_\mu^{ab} =0   \;, \label{brst3}
\end{eqnarray}
with 
\begin{equation}
s_{\gamma^2}^2=0\;, \qquad
s_{\gamma^2} S_{GZ} = 0 \;. \label{gzinv}
\end{equation}
The  operator $s_{\gamma^2}$ depends explicitly on the non-perturbative Gribov parameter $\gamma^2$. As such, it is a non-perturbative extension of the usual BRST operator of the Faddeev-Popov theory, to which it reduces  in the limit $\gamma^2 \rightarrow 0$. \\\\Before discussing the generalization of this construction to the case of the maximal Abelian gauge, it is worth spending a few words on the nature of the non-local variable $A^h_\mu$ of expression  \eqref{hhh3g}. As already pointed out, the field $A^h_\mu$ is left invariant order by order by infinitesimal gauge transformations. As a consequence, it is invariant under the action of the BRST operator $s_{\gamma^2}$, {\it i.e.}
\begin{equation}
s_{\gamma^2} A^h_\mu = 0 \;. \label{invah}
\end{equation}
Equation \eqref{invah} might give the impression that the gauge field $A^h_\mu$ could create an invariant physical state carrying a color index. Nevertheless, in the present case, this possibility is ruled out by the presence of the Gribov parameter $\gamma^2$, which acts as a confining parameter for gluons in the non-perturbative infrared region. In fact, a look at the tree-level two-point correlation function $\langle A^h_\mu(k) A^h_\nu(-k) \rangle_{GZ}$ stemming from the Gribov-Zwanziger action \eqref{gzh3} learns that
\begin{equation}
\langle A^h_\mu(k) A^h_\nu(-k) \rangle_{GZ} = \frac{k^2}{k^4+2g^2N\gamma^4} \left( \delta_{\mu \nu} - \frac{k_\mu k_\nu}{k^2} \right)   \;. \label{gzprop}
\end{equation}
Expression \eqref{gzprop} displays complex poles, a feature which prevents a particle interpretation. In other words, even if the field $A^h_\mu$ is left invariant by the operator $s_{\gamma^2}$, it cannot be associated to a physical state due to the confining character of the Gribov horizon, encoded in the Gribov mass parameter $\gamma^2$. These considerations can be easily extended to the case of the refined-Gribov-Zwanziger theory \cite{Dudal:2007cw,Dudal:2008sp,Dudal:2011gd} as well as to the maximal Abelian gauge. Of course, the field $A^h_\mu$ could be introduced also in the case of a non-confining theory, like the electroweak  theory $SU(2) \times U(1)$, see  \cite{Capri:2012ah}. In this case, the use of the variable $A^h_\mu$ leads to a gauge invariant description of the physical vector bosons $(W^{+}, W^{-}, Z^{0})$, as discussed in details in  \cite{Lavelle:1995ty}. \\\\The introduction of the field $A^h_\mu$ and of the operator $s_{\gamma^2}$ allows us to clarify better the physical meaning of the Gribov parameter $\gamma^2$. As it is apparent from expressions \eqref{gzh2}, \eqref{gzh3}, the horizon function $\gamma^4 H(A^h)$ acquires now the meaning of a quantity invariant under the BRST operator $s_{\gamma^2}$. This is an interesting feature, enabling us to state that the Gribov parameter $\gamma^2$ is not akin to an unphysical gauge parameter. In fact, from 
\begin{equation}
s_{\gamma^2}  H(A^h) = 0\;, \label{invH} 
\end{equation}
it immediately follows that 
\begin{equation}
\frac{\partial S_{GZ}}{\partial \gamma^2} \neq s_{\gamma^2} (....)    \;, \label{nt}
\end{equation}
meaning that $\gamma^2$ is a physical parameter of the theory. As such, it will enter in the expressions of the gauge invariant correlation functions, deeply modifying their behavior in the infrared region, as explicitly reported, for example, in the evaluation of the spectral densities for the glueball spectrum \cite{Dudal:2010cd,Dudal:2013wja}.

\section{Generalization to the maximal Abelian gauge} 
In order to generalize the previous set up to the maximal Abelian gauge, let us start by fixing the notation and by reminding a few properties of this gauge. 

\subsection{The maximal Abelian gauge fixing and the horizon function} 

We shall consider, for simplicity, the gauge group $SU(2)$, whose generators $T^a$, $(a=1,2,3)$, 
\begin{equation}
\left[ T^a,T^b\right] =i\varepsilon ^{abc}T^c\,\,  \label{la}
\end{equation}
are chosen to be hermitian and to obey the orthonormality condition $\,\mathrm{Tr}\left( T^aT^b\right) =\delta ^{ab}$. Following  \cite{'tHooft:1981ht,Kronfeld:1987vd,Kronfeld:1987ri}, we decompose  the gauge field ${A}_\mu $  into off-diagonal and diagonal components 
\begin{equation}
{A}_\mu ={A}_\mu ^aT^a=A_\mu ^{\alpha} T^{\alpha}+A_\mu T^{\,3},  \label{cd}
\end{equation}
where $\alpha=1,2$ and $T^3$ is the diagonal generator of the Cartan subgroup of $SU(2)$. Analogously, decomposing the field strength, we obtain
\begin{equation}
{F}_{\mu \nu }={F}_{\mu \nu }^aT^a=F_{\mu \nu }^{\alpha}T^{\alpha}+F_{\mu
\nu }T^{\,3},  \label{fs}
\end{equation}
with the off-diagonal and diagonal components given, respectively, by
\begin{eqnarray}
F_{\mu \nu }^{\alpha}  &=&D_\mu ^{\alpha \beta}A_\nu ^{\beta} -D_\nu ^{\alpha \beta}A_\mu^{\beta}\,,  \nonumber \\
F_{\mu \nu } &=&\partial _\mu A_\nu -\partial _\nu A_\mu +g \varepsilon
^{\alpha \beta}A_\mu^{\alpha} A_\nu^{\beta}\,,  \label{fscomp}
\end{eqnarray}
where the covariant derivative $D_\mu ^{\alpha \beta}$ is defined with respect to the
diagonal component $A_\mu $
\begin{equation}
D_\mu ^{\alpha \beta}\equiv \partial _\mu \delta ^{\alpha \beta}-g\varepsilon^{\alpha \beta}A_\mu
\,\,\,\,\,\,,\,\,\,\varepsilon^{\alpha \beta}\equiv \varepsilon^{\alpha \beta 3}\,\,\,.
\label{cder}
\end{equation}
Also, the Yang-Mills action 
\begin{equation}
S_{YM}=\int d^4x\; \frac{1}{4} \left(F_{\mu\nu}^{\alpha} F_{\mu\nu}^{\alpha}+F_{\mu\nu}F_{\mu\nu}\right)  \;, \label{ym}
\end{equation}
is left invariant by the gauge transformations
\begin{eqnarray}
\delta A_{\mu }^{\alpha} &=&-D_{\mu }^{\alpha \beta}{\omega }^{\beta}-g\varepsilon^{\alpha \beta}A_{\mu
}^{\beta}\omega \;,  \nonumber \\
\delta A_{\mu } &=&-\partial _{\mu }{\omega }-g\varepsilon^{\alpha \beta}A_{\mu
}^{\alpha}\omega^{\beta}\;.  \label{gauge}
\end{eqnarray}
The maximal Abelian gauge condition amounts to impose that the off-diagonal
components $A_{\mu }^{\alpha}$ of the gauge field obey the following non-linear condition
\begin{equation}
D_{\mu }^{\alpha \beta}A_{\mu }^{\beta}=0\;.  \label{offgauge}
\end{equation}
Moreover, as it is apparent from the presence of the covariant derivative $D_{\mu }^{\alpha \beta}$, equation (\ref{offgauge}) allows for a residual local $U(1)$
invariance corresponding to the diagonal subgroup of $SU(2)$. This
additional invariance has to be fixed by means of a further gauge condition
on the diagonal component $A_{\mu }$, which is usually chosen to be of the
Landau type, namely
\begin{equation}
\partial _{\mu }A_{\mu }=0\;.  \label{dgauge}
\end{equation}
The Faddeev-Popov operator, $\mathcal{M}^{\alpha \beta}$, corresponding to the gauge
condition (\ref{offgauge}) is easily derived, being given by  
\begin{equation}
\mathcal{M}^{\alpha \beta}=-D_{\mu }^{\alpha \delta}D_{\mu }^{\delta \beta}-g^{2}\varepsilon
^{\alpha \sigma}\varepsilon ^{\beta \delta }A_{\mu }^{\sigma}A_{\mu }^{\delta}\;.  \label{offop}
\end{equation}
In order to construct the Faddeev-Popov action corresponding to the gauge conditions \eqref{offgauge},\eqref{dgauge}, we proceed by introducing the standard nilpotent BRST transformations 
\begin{eqnarray}
sA^{\alpha}_{\mu}& = & -(D^{\alpha \beta }_{\mu}c^{\beta}+g\varepsilon^{\alpha \beta}A_{\mu}^{\beta} c)\,,\qquad sA_{\mu}=-(\partial_{\mu}c+g\varepsilon^{\alpha \beta}A_{\mu}^{\alpha} c^{\beta})\,,\nonumber\\ 
sc^{\alpha}& = & g\varepsilon^{\alpha \beta }c^{\beta} c\,,\qquad sc=\frac{g}{2}\varepsilon^{\alpha \beta }c^{\alpha}c^{\beta} \,, \nonumber \\
s\bar{c}^{\alpha}& = & b^{\alpha}\,,\qquad s\bar{c}=b\,,\qquad
sb^{\alpha}=sb=0 \;, 
\label{stransf}
\end{eqnarray}
where $({\bar c}^{\alpha}, {\bar c}, c^{\alpha}, c)$ are the Faddeev-Popov ghosts and $(b^{\alpha},b)$ are the Lagrange multipliers implementing the gauge conditions \eqref{offgauge},\eqref{dgauge}. Further, we introduce the 
$s$-exact gauge fixing term
\begin{eqnarray}
S_{MAG}&=&s\int d^4x\left( \bar{c}^{\alpha}D^{\alpha \beta}_{\mu}A^{\beta}_{\mu}+\bar{c}\partial_\mu A_{\mu}\right)\nonumber\\
&=&\int d^4x\left( b^{\alpha} D_{\mu}^{\alpha \beta}A_{\mu}^{\beta} -\bar{c}^{\alpha} \mathcal{M}^{\alpha \beta}c^{\beta} 
+g\varepsilon^{\alpha \beta}\bar{c}^{\alpha }  (D_{\mu}^{\beta \delta}A_{\mu}^{\delta}) c +b\partial_{\mu}A_{\mu}
+\bar{c}\,\partial_{\mu}\left(\partial_{\mu}c+g\varepsilon^{\alpha \beta}A_{\mu}^{\alpha} c^{\beta} \right)\right)  \;, \nonumber \\ \label{smag}
\end{eqnarray}
where $\mathcal{M}^{ab}$ stands for the Faddeev-Popov operator \eqref{offop}. 
It is apparent that the Faddeev-Popov action of the maximal Abelian gauge 
\begin{equation}
S_{MAG}^{FP} = S_{YM} + S_{MAG}  \;, \label{act1}
\end{equation}
turns out to be BRST invariant, {\it i.e.} 
\begin{equation} 
s S_{MAG}^{FP} =0 \;. \label{invfp} 
\end{equation}
As any other covariant gauge, also the maximal Abelian gauge is plagued by the existence of Gribov copies, see refs.\cite{Bruckmann:2000xd,Guimaraes:2011sf,Capri:2013vka} for explicit examples of zero modes of the Faddeev-Popov operator \eqref{offop}.  By restricting the integration in the functional  integral to the region where the Faddeev-Popov operator $\mathcal{M}^{\alpha \beta}$ is strictly positive, {\it i.e.} $\mathcal{M}^{\alpha \beta}>0$, a large number of copies could be eliminated, as proven in \cite{Capri:2005tj,Capri:2008vk}. Furthermore, in complete analogy with the case of the Landau gauge, this restriction is implemented by adding to the original Faddeev-Popov action, eq.\eqref{act1},  a non-local horizon term which, in the case of the maximal Abelian gauge, turns out to be given by the expression \cite{Capri:2005tj,Capri:2006cz,Capri:2008ak,Capri:2008vk}
\begin{equation}
H_{MAG}(A)= g^2\int d^4x\;d^4y\;A_{\mu}(x)\varepsilon^{\alpha \beta}\left(\mathcal{M}^{-1}\right)^{\alpha \delta}(x,y)\varepsilon^{\delta \beta}A_{\mu}(y)\,.
\label{H_MAG}
\end{equation}
Therefore, for the analogous of the Gribov-Zwanziger action in the maximal Abelian gauge, we have 
\begin{equation}
S^{GZ}_{MAG} = S_{MAG}^{FP} + \gamma^4 H_{MAG}(A) \;, \label{gzmag}
\end{equation}
where $\gamma^2$ stands for the Gribov parameter of the maximal Abelian gauge. Proceeding as in the case of the Landau gauge, expression \eqref{gzmag} can be cast in local form by introducing a pair of auxiliary bosonic fields, $(\bar{\phi}^{\alpha \beta}_{\mu}, {\phi}^{\alpha \beta}_{\mu})$, and a pair of auxiliary fermonic fields, $(\bar{\omega}^{\alpha \beta}_{\mu}, {\omega}^{\alpha \beta}_{\mu})$, namely 
\begin{equation} 
S^{GZ}_{MAG} = S_{MAG}^{FP} + \int d^4x\left(\bar{\phi}^{\alpha \beta}_{\mu}\mathcal{M}^{\alpha \delta}\phi^{\delta \beta}_{\mu}
-\bar{\omega}^{\alpha \beta}_\mu \mathcal{M}^{\alpha \delta}\omega^{\delta \beta }_\mu + g\gamma^2 \varepsilon^{\alpha \beta}\left(\phi^{\alpha \beta }_\mu -\bar{\phi}^{\alpha \beta}_\mu \right)A_{\mu}  \right)  \;.   \label{locgzmag}
\end{equation} 
As shown in \cite{Capri:2005tj,Capri:2006cz,Capri:2008ak,Capri:2008vk}, the action $S^{GZ}_{MAG}$ enables us to implement the restriction in the functional integral to the Gribov region $\Omega_{MAG}$ of the maximal Abelian gauge, defined as 
\begin{equation}
\Omega_{MAG} = \left\{ \; A^\alpha_\mu\,, A_{\mu}\;\;\left| \;  \;  \partial_\mu A_\mu =0,  \; D^{\alpha \beta}_\mu A^{\beta}_\mu=0,  \;    {\mathcal M}^{\alpha \beta}(A) =  -\left( D_{\mu }^{\alpha \delta}D_{\mu }^{\delta \beta}+g^{2}\varepsilon
^{\alpha \sigma}\varepsilon ^{\beta \delta }A_{\mu }^{\sigma}A_{\mu }^{\delta} \right)> 0  \;\right. \right\}    \;. \label{om}
\end{equation}
Although the understanding of the Gribov issue in the maximal Abelian gauge cannot yet be compared to that reached in the Landau gauge, a few properties of the region $\Omega_{MAG}$ have been already obtained. In particular, in  \cite{Capri:2008vk}, it has been established that $\Omega_{MAG}$ is unbounded along the diagonal directions in field space. This feature seems to be consistent with the aforementioned Abelian dominance hypothesis, according to which the diagonal  configurations, corresponding to the Abelian Cartan subgroup, should be the dominant configurations in the infrared. Moreover, in  \cite{Greensite:2004ke}, it has been shown that when an Abelian configuration is gauge-transformed to the Landau gauge, it is mapped into a point of the boundary of the Gribov region $\Omega$ of the Landau gauge, eq.\eqref{om}, {\it i.e.} into a point of the Gribov horizon\footnote{See Sect.V of  \cite{Greensite:2004ke}.}. These features give further support to the restriction of the domain of integration to the  region $\Omega_{MAG}$. \\\\We can now address the issue of the  existence of a nilpotent non-perturbative BRST symmetry for the action \eqref{locgzmag}.

\subsection{The non-perturbative BRST symmetry for the maximal Abelian gauge} 
In order to construct a non-perturbative BRST symmetry for the action \eqref{locgzmag}, we proceed as in the case of the Landau gauge. Making  use of the invariant field $A^h_\mu$, eq.\eqref{hhh3g}, we can rewrite expression  \eqref{H_MAG} as 
\begin{eqnarray} 
H_{MAG}(A) &= & H_{MAG}(A^h) - {\mathcal F}(A) \partial A -  {\mathcal F}^{\alpha}(A) \partial A^{\alpha} \nonumber \\
& = &  H_{MAG}(A^h) - \left( - \partial_\mu {\mathcal F}(A) + g \varepsilon^{\alpha \beta}  {\mathcal F}^{\alpha}(A) A^\beta_\mu \right) A_\mu - {\mathcal F}^{\alpha}(A) D^{\alpha \beta}_\mu A^\beta_\mu 
\end{eqnarray} 
where, as in the previous section,  $ {\mathcal F}(A) \partial A $, $ {\mathcal F}^{\alpha}(A) \partial A^{\alpha}$  stand for a  short-hand notation, {\it i.e.}   ${\mathcal F}(A) (\partial A)= \int d^4x d^4y {\mathcal F}(x,y) (\partial A)_y$ and  ${\mathcal F}^{\alpha}(A) (\partial A^{\alpha})= \int d^4x d^4y {\mathcal F}^{\alpha}(x,y) (\partial A^\alpha)_y$. Introducing as before the redefined Lagrange multipliers $b^h, b^{h,\alpha}$ 
\begin{eqnarray} 
b^h & = & b -\gamma^4 {\mathcal F}(A) + \gamma^4 \int_{-\infty}^{x} dy_\mu \left(  g \varepsilon^{\alpha \beta}  {\mathcal F}^{\alpha}(A) A^\beta_\mu \right)_y \nonumber \\
b^{h,\alpha} & = & b^{\alpha} - \gamma^4 {\mathcal F}^{\alpha}(A)   \;, \label{redb}
\end{eqnarray}
we can rewrite the action \eqref{gzmag} as 
\begin{equation}
S^{GZ}_{MAG} = S_{YM} + S_{MAG}(b^h, b^{h,\alpha})  + \gamma^4 H_{MAG}(A^h) \;, \label{gzmag1}
\end{equation}
with
\begin{equation}
S_{MAG}(b^h, b^{h,\alpha}) 
=\int d^4x\left( b^{h, \alpha} D_{\mu}^{\alpha \beta}A_{\mu}^{\beta} -\bar{c}^{\alpha} \mathcal{M}^{\alpha \beta}c^{\beta} 
+g\varepsilon^{\alpha \beta}\bar{c}^{\alpha} (D_{\mu}^{\beta \delta}A_{\mu}^{\delta}) c +b^h\partial_{\mu}A_{\mu}
+\bar{c}\,\partial_{\mu}\left(\partial_{\mu}c+g\varepsilon^{\alpha \beta}A_{\mu}^{\alpha} c^{\beta} \right)\right)  \label{smag1}
\end{equation}
Notice that equations \eqref{redb}  correspond to a change of variables in the functional integral in the $b$-sector of the theory, the corresponding Jacobian being the unity. \\\\Introducing thus the auxiliary fileds $(\bar{\phi}, {\phi}, \bar{\omega}, {\omega})$, we obtain 
\begin{eqnarray}
S^{GZ}_{MAG} & = & S_{YM}  +  \int d^4x\left(\bar{\phi}^{\alpha \beta}_{\mu}\mathcal{M}^{\alpha \delta}(A^h) \phi^{\delta \beta}_{\mu}
-\bar{\omega}^{\alpha \beta}_\mu \mathcal{M}^{\alpha \delta}(A^h) \omega^{\delta \beta }_\mu + g\gamma^2 \varepsilon^{\alpha \beta}\left(\phi^{\alpha \beta }_\mu -\bar{\phi}^{\alpha \beta}_\mu \right)A^{h,3}_{\mu}  \right)   \nonumber \\
&+& \int d^4x\left( b^{h, \alpha} D_{\mu}^{\alpha \beta}A_{\mu}^{\beta} -\bar{c}^{\alpha} \mathcal{M}^{\alpha \beta}c^{\beta} 
+g\varepsilon^{\alpha \beta}\bar{c}^{\alpha} (D_{\mu}^{\beta \delta}A_{\mu}^{\delta}) c +b^h\partial_{\mu}A_{\mu}
+\bar{c}\,\partial_{\mu}\left(\partial_{\mu}c+g\varepsilon^{\alpha \beta}A_{\mu}^{\alpha} c^{\beta} \right)\right)  \;. \nonumber \\
\label{fmaggz1} 
\end{eqnarray} 
As in the case of the Landau gauge, expression \eqref{fmaggz1} enjoys a non-perturbative exact nilpotent BRST symmetry, namely 
\begin{eqnarray}
s_{\gamma^2} A^{\alpha}_{\mu}& = & -(D^{\alpha \beta }_{\mu}c^{\beta}+g\varepsilon^{\alpha \beta}A_{\mu}^{\beta} c)\,,\qquad s_{\gamma^2} A_{\mu}=-(\partial_{\mu}c+g\varepsilon^{\alpha \beta}A_{\mu}^{\alpha} c^{\beta})\,,\nonumber\\ 
s_{\gamma^2}c^{\alpha}& = & g\varepsilon^{\alpha \beta }c^{\beta} c\,,\qquad s_{\gamma^2}c=\frac{g}{2}\varepsilon^{\alpha \beta }c^{\alpha}c^{\beta} \,, \nonumber \\
s_{\gamma^2} \bar{c}^{\alpha}& = & b^{h, \alpha}\,,\qquad s_{\gamma^2} \bar{c}=b^h \,, \nonumber \\
s_{\gamma^2} b^{h,\alpha}& = & s_{\gamma^2} b=0 \;, \nonumber \\
s_{\gamma^2} \phi^{\alpha \beta}_{\mu}  &=&  \omega^{\alpha \beta}_{\mu} \;, \qquad s_{\gamma^2} \omega^{\alpha \beta}_{\mu} = 0 \;, \nonumber \\
s_{\gamma^2} {\bar \omega}^{\alpha \beta}_{\mu} & = & \bar{\phi}^{\alpha \beta}_{\mu} + g \gamma^2 \varepsilon^{\delta \beta} A^{h,3}_\mu \left( {\mathcal M}^{-1} (A^h) \right)^{\delta \alpha}  \;, \qquad s_{\gamma^2} \bar{\phi}^{\alpha \beta}_{\mu}=0 \;, 
\label{npbrstmag}
\end{eqnarray}
and 
\begin{equation}
s_{\gamma^2} s_{\gamma^2} =0 \;, \qquad s_{\gamma^2}S^{GZ}_{MAG} =0 \;.    \label{exmag}
\end{equation}
Let us end this section by mentioning that, although we have considered the gauge group $SU(2)$, the above construction can be straightforwardly generalized to the gauge group $SU(N)$ by making use of the results of \cite{Capri:2010an}.

\section{The Refined Gribov-Zwanziger action in the maximal Abelian gauge} 
As already mentioned, the Gribov-Zwanziger action in the Landau gauge, eq.\eqref{gzh3}, generalizes to its refined version \cite{Dudal:2007cw,Dudal:2008sp,Dudal:2011gd}. The origin of the refined version stems for the observation that non-perturbative condensates of dimensions two, $\langle A^{h,a}_\mu A^{h,a}_\mu \rangle $ and $\langle {\bar \phi}^{ab}_\mu \phi^{ab}_\mu - {\bar  \omega}^{ab}_\mu \omega^{ab}_\mu \rangle$, are dynamically generated for non-vanishing Gribov parameter $\gamma^2$. The effective action taking into account the existence of these condensates is called the refined Gribov-Zwanziger action \cite{Dudal:2007cw,Dudal:2008sp,Dudal:2011gd}. So far, the prediction of the refined action on the infrared behaviour of the gluon and ghost propagators are in very good agreement with the most recent lattice data, see  \cite{Dudal:2010tf,Cucchieri:2011ig}. \\\\A refined version of the action \eqref{fmaggz1}  in the maximal Abelian gauge has also been constructed in  \cite{Capri:2008ak}, where the analogous of the dimension two condensates of the Landau gauge have been introduced. More precisely, in the case of the maximal Abelian gauge, due to the splitting of the gauge field into diagonal and off-diagonal components, we have the following dimension two condensates: $\langle A^{h,\alpha}_\mu A^{h,\alpha}_\mu \rangle $, $\langle A^{h,3}_\mu A^{h,3}_\mu \rangle $ and $\langle {\bar \phi}^{\alpha \beta}_\mu \phi^{\alpha \beta}_\mu - {\bar  \omega}^{\alpha \beta}_\mu \omega^{\alpha \beta}_\mu \rangle$, playing a different role at the dynamical level. \\\\In particular, the condensate $\langle A^{h,\alpha}_\mu A^{h,\alpha}_\mu \rangle $ provides a dynamical Yukawa type mass for the off-diagonal components of the gluon field \cite{Dudal:2004rx,Kondo:2003uq}. This gives support to the Abelian dominance hypothesis, according to which the off-diagonal components should decouple at very low energy scales \cite{'tHooft:1981ht,Ezawa:bf}\footnote{As a consequence of the existence of the condensate $\langle A^{h,\alpha}_\mu A^{h,\alpha}_\mu \rangle $, the tree level off-diagonal propagator takes the form 
\begin{equation} 
\langle A^{\alpha}_\mu(k) A^{\beta}_\nu(-k) \rangle = \delta^{\alpha\beta}\left( \delta_{\mu\nu} - \frac{k_\mu k_\nu}{k^2} \right) \frac{1}{k^2+m^2_{\rm off}} \;, \label{off}
\end{equation}
where the dynamical mass parameter $m^2_{\rm off}$ is related to  $\langle A^{h,\alpha}_\mu A^{h,\alpha}_\mu \rangle $. Numerical lattice simulations \cite{Ezawa:bf,Sasaki:1998ww,Sasaki:1998th,Suzuki:1995va,Suzuki:1989gp,Shiba:1994ab,Stack:1994wm,Hioki:1991ai,Miyamura:1995xn,Amemiya:1998jz,Bornyakov:2003ee,Gongyo:2013sha,Sakumichi:2014xpa,Mendes:2007vv,Mendes:2008vf} have given evidence that the off-diagonal mass $m^2_{\rm off}$ is large enough so as the off-diagonal gluon components are more suppressed than the corresponding diagonal components at very low energy scales.}. \\\\On the other hand, the condensates $\langle A^{h,3}_\mu A^{h,3}_\mu \rangle $ and $\langle {\bar \phi}^{\alpha \beta}_\mu \phi^{\alpha \beta}_\mu - {\bar  \omega}^{\alpha \beta}_\mu \omega^{\alpha \beta}_\mu \rangle$ determine the infrared behaviour of the diagonal gluon propagator. Following \cite{Capri:2008ak}, for the refined action which takes into account the dimension two condensates in the maximal Abelian gauge,  we have 
\begin{equation}
S^{RGZ}_{MAG}  =  S^{GZ}_{MAG} + \frac{m^2_{\rm off}}{2} \int d^4x\, A^{h,\alpha}_\mu A^{h,\alpha}_\mu   + \int d^4x \left( \mu^2 ( {\bar \phi}^{\alpha \beta}_\mu \phi^{\alpha \beta}_\mu - {\bar  \omega}^{\alpha \beta}_\mu \omega^{\alpha \beta}_\mu ) + \frac{m^2_{\rm diag}}{2}  A^{h,3}_\mu A^{h,3}_\mu \right)    \label{refmag} \;, 
\end{equation} 
with $S^{GZ}_{MAG}$  given in expression \eqref{fmaggz1}. \\\\The parameters $(m^2_{\rm off}, \mu^2, m^2_{\rm diag}) $ have a dynamical origin, encoding the presence of the dimension two condensates $\langle A^{h,\alpha}_\mu A^{h,\alpha}_\mu \rangle $, $\langle A^{h,3}_\mu A^{h,3}_\mu \rangle $ and $\langle {\bar \phi}^{\alpha \beta}_\mu \phi^{\alpha \beta}_\mu - {\bar  \omega}^{\alpha \beta}_\mu \omega^{\alpha \beta}_\mu \rangle$. 
The non-perturbative BRST nilpotent symmetry introduced in the previous section  generalizes almost immediately to the case of the refined action \eqref{refmag}. The only difference with respect to equations \eqref{npbrstmag} is given by the transformation of the field ${\bar \omega}^{\alpha \beta}_{\mu}$, which now reads 
\begin{equation}
s_{\gamma^2} {\bar \omega}^{\alpha \beta}_{\mu}  =  \bar{\phi}^{\alpha \beta}_{\mu} + g \gamma^2 \varepsilon^{\delta \beta} A^{h,3}_\mu \left( \left[{\mathcal M}(A^h)+\mu^2\bold{1} \right]^{-1} \right)^{\delta \alpha}    \label{newnpbrst} \;. 
\end{equation}
Again, one checks that 
\begin{equation}
s_{\gamma^2} s_{\gamma^2} =0 \;, \qquad s_{\gamma^2}S^{RGZ}_{MAG} =0 \;.    \label{exfrgz}
\end{equation}
In view of the analysis of the diagonal gluon propagator, it is instructive to present the explicit first-order evaluation of the condensates $\langle A^{h,3}_\mu A^{h,3}_\mu \rangle $ and $\langle {\bar \phi}^{\alpha \beta}_\mu \phi^{\alpha \beta}_\mu - {\bar  \omega}^{\alpha \beta}_\mu \omega^{\alpha \beta}_\mu \rangle$. \\\\At the first-order only the quadratic part of the action \eqref{fmaggz1} is needed, namely 
\begin{eqnarray} 
(S^{GZ}_{MAG})_{\rm quad} & = & \int d^4x \left(  \frac{1}{2} A^{\alpha}_\mu (-\delta_{\mu\nu} \partial^2 +\partial_\mu \partial_\nu) A^{\alpha}_\nu + \frac{1}{2} A_\mu (-\delta_{\mu\nu} \partial^2 +\partial_\mu \partial_\nu) A_\nu  \right) \nonumber \\
& + & \int d^4x \left( b^{\alpha} \partial_\mu A^{\alpha}_\mu + {\bar c}^{\alpha} \partial^2 c^{\alpha} + b \partial_\mu A_\mu + {\bar c} \partial^2 c \right) \nonumber \\
& + & \int d^4x \left(  {\bar \phi}^{\alpha \beta}_\mu (-\partial^2) {\phi}^{\alpha \beta}_\mu  - {\bar \omega}^{\alpha \beta}_\mu (-\partial^2) {\omega}^{\alpha \beta}_\mu
+ g\gamma^2 \varepsilon^{\alpha \beta}\left(\phi^{\alpha \beta }_\mu -\bar{\phi}^{\alpha \beta}_\mu \right)A_{\mu}  \right) \;. \label{quadr}
\end{eqnarray} 
Further, we  introduce the operators  $\int d^4x A^{h,3}_\mu A^{h,3}_\mu $ and  $\int d^4x ( {\bar \phi}^{\alpha \beta}_\mu \phi^{\alpha \beta}_\mu - {\bar  \omega}^{\alpha \beta}_\mu \omega^{\alpha \beta}_\mu)$ in the action by coupling them to two constant sources $J$ and $\sigma$, and we define the vacuum functional ${\cal E}(\sigma,J)$ defined by 
\begin{equation}
\mathrm{e}^{-V{\cal E}(\sigma,J)}=\int \left[D\Phi\right]\mathrm{e}^{-(S^{GZ}_{MAG})_{\rm quad} - J\int d^4x ( {\bar \phi}^{\alpha \beta}_\mu \phi^{\alpha \beta}_\mu - {\bar  \omega}^{\alpha \beta}_\mu \omega^{\alpha \beta}_\mu)-\sigma\int d^4x~A_{\mu}\left(\delta_{\mu\nu}-\frac{\partial_{\mu}\partial_{\nu}}{\partial^{2}}\right)A_{\nu}}\,,
\label{cond2}
\end{equation}
where we used the fact that, at the first-order, 
\begin{equation}
\int d^4x \;A^{h,3}_\mu A^{h,3}_\mu =  \int d^4x~A_{\mu}\left(\delta_{\mu\nu}-\frac{\partial_{\mu}\partial_{\nu}}{\partial^{2}}\right)A_{\nu}  \;. \label{fo}
\end{equation} 
It is easy  to check that the  condensates  $\langle A^{h,3}_\mu A^{h,3}_\mu \rangle $ and $\langle {\bar \phi}^{\alpha \beta}_\mu \phi^{\alpha \beta}_\mu - {\bar  \omega}^{\alpha \beta}_\mu \omega^{\alpha \beta}_\mu \rangle$ are obtained by differentiating ${\cal E}(\sigma,J)$ with respect to the sources $(J,\sigma)$, which are set to zero at the end, {\it i.e.}
\begin{eqnarray}
\langle {\bar \phi}^{\alpha \beta}_\mu \phi^{\alpha \beta}_\mu - {\bar  \omega}^{\alpha \beta}_\mu \omega^{\alpha \beta}_\mu \rangle &=& \frac{\partial {\cal E}(\sigma,J)}{\partial J}\Big|_{J=\sigma=0} \nonumber  \\
\langle A^{h,3}_\mu A^{h,3}_\mu \rangle &=& \frac{\partial {\cal E}(\sigma,J)}{\partial \sigma}\Big|_{J=\sigma=0}\,.
\label{cond3}
\end{eqnarray}
Employing dimensional regularization, a direct computation shows that
\begin{equation}
{\cal E}(\sigma,J)=\frac{(D-1)}{2}\int \frac{d^Dk}{(2\pi)^D}~\mathrm{ln}\left(k^2+\frac{4\gamma^4g^2}{k^2+J}+2\sigma \right) + \;\;{\rm terms\; independ.\;from\; ({\it J},\sigma) } \,. 
\label{cond6}
\end{equation}

\noindent Eq.(\ref{cond3}) and (\ref{cond6}) give thus

\begin{equation}
\langle {\bar \phi}^{\alpha \beta}_\mu \phi^{\alpha \beta}_\mu - {\bar  \omega}^{\alpha \beta}_\mu \omega^{\alpha \beta}_\mu \rangle_{\mbox{\tiny first-order}}  = -2 \gamma^4g^2 (D-1)\int \frac{d^Dk}{(2\pi)^D}\frac{1}{k^2}\frac{1}{(k^4+4g^2\gamma^4)}
\label{cond7}
\end{equation}

\noindent and

\begin{equation}
\langle A^{h,3}_\mu A^{h,3}_\mu \rangle_{\mbox{\tiny first-order}} = -4 g^2 \gamma^4(D-1)\int\frac{d^Dk}{(2\pi)^D}\frac{1}{k^2}\frac{1}{(k^4+4g^2\gamma^4)}\,.
\label{cond8}
\end{equation}
Eq.(\ref{cond7}) and eq.(\ref{cond8}) show that, already at first-order, both condensates  $\langle   A^{h,3}_\mu A^{h,3}_\mu \rangle$ and $\langle {\bar \phi}^{\alpha \beta}_\mu \phi^{\alpha \beta}_\mu - {\bar  \omega}^{\alpha \beta}_\mu \omega^{\alpha \beta}_\mu \rangle$ are non-vanishing and proportional to the Gribov parameter $\gamma^2$. Notice also that both integrals in eqs.(\ref{cond7}), (\ref{cond8})  are perfectly convergent in the ultraviolet region by power counting. We see thus that, in perfect analogy with the case of the Landau gauge \cite{Dudal:2007cw,Dudal:2008sp,Dudal:2011gd}, dimension two condensates are automatically generated by the presence of the Gribov parameter $\gamma^2$. \\\\Having introduced the refined action, eq.\eqref{refmag}, it is helpful to add a few remarks on its meaning. To that aim, we integrate out the auxiliary fields  $({\bar \phi}^{\alpha \beta}_\mu, \phi^{\alpha \beta}_\mu,  {\bar  \omega}^{\alpha \beta}_\mu \omega^{\alpha \beta}_\mu)$,  obtaining 
\begin{eqnarray} 
S^{RGZ}_{MAG}  & = & S_{YM} +   g^2\gamma^4 \int d^4x\;d^4y\;A^{h,3}_{\mu}(x)\varepsilon^{\alpha \beta}\left( \left[{\mathcal M}(A^h)+\mu^2\bold{1} \right]^{-1} \right)^{\alpha \delta}_{x,y}\; \varepsilon^{\delta \beta}A^{h,3}_{\mu}(y) \nonumber \\
&+&  \frac{m^2_{\rm off}}{2} \int d^4x\, A^{h,\alpha}_\mu A^{h,\alpha}_\mu   + \int d^4x \left( \mu^2 ( {\bar \phi}^{\alpha \beta}_\mu \phi^{\alpha \beta}_\mu - {\bar  \omega}^{\alpha \beta}_\mu \omega^{\alpha \beta}_\mu ) + \frac{m^2_{\rm diag}}{2}  A^{h,3}_\mu A^{h,3}_\mu \right)   \nonumber \\
&+& \int d^4x\left( b^{h, \alpha} D_{\mu}^{\alpha \beta}A_{\mu}^{\beta} -\bar{c}^{\alpha} \mathcal{M}^{\alpha \beta}c^{\beta} 
+g\varepsilon^{\alpha \beta}\bar{c}^{\alpha} (D_{\mu}^{\beta \delta}A_{\mu}^{\delta}) c +b^h\partial_{\mu}A_{\mu}
+\bar{c}\,\partial_{\mu}\left(\partial_{\mu}c+g\varepsilon^{\alpha \beta}A_{\mu}^{\alpha} c^{\beta} \right)\right)  \;, \nonumber \\
 \label{magaa}  
\end{eqnarray} 
from which one sees that the starting horizon function \eqref{H_MAG} gets replaced by the expression
\begin{equation}
H_{MAG}(A^h) \Rightarrow  g^2\gamma^4 \int A^{h} \left( \left[{\mathcal M}(A^h)+\mu^2\bold{1} \right]^{-1} \right) \; A^{h}   \;. \label{ha1}
\end{equation}
Nevertheless, one also observes that the Faddeev-Popov operator $\mathcal{M}^{\alpha \beta}(A)$ entering the Faddeev-Popov ghost sector of expression \eqref{magaa} is left unchanged, being given by the term $\bar{c}^{\alpha} \mathcal{M}^{\alpha \beta}(A)c^{\beta} $. Thus, the question which naturally arises is: how the properties of the Faddeev-Popov ghost correlation function $\langle {\bar c}^{\alpha} c^\beta \rangle = \left(  \mathcal{M}^{-1} \right)^{\alpha \beta}$ are changed when the gluon sector has been modified according  to eq.\eqref{ha1}?  To answer this question we shall partly refer to the case of the Landau gauge, where the same situation is found. First, we underline that the parameter $\mu^2$ appearing in the refined action \eqref{refmag} is not free. In fact, from expression \eqref{cond7}, one realizes that the dimension two condensate $\langle {\bar \phi} \phi - {\bar  \omega} \omega \rangle$ is proportional  to the Gribov parameter $\gamma^2$. This  means that the mass parameter $\mu^2$ has a dynamical origin, being generated by the Gribov parameter itself. Furthermore, as shown in great detail in \cite{Dudal:2008sp} in the case of the Landau gauge, where an explicit check of the no-pole Gribov condition has been worked out, with the inclusion of the dimension two condensate $\langle {\bar \phi} \phi - {\bar  \omega}  \omega \rangle$ the Faddeev-Popov ghost correlation function $\langle {\bar c} c \rangle = \left(  \mathcal{M}^{-1} \right)$  retains the fundamental property of being positive.  This means that with the inclusion of the condensate one remains inside the Gribov region $\Omega$, {\it i.e.} the Gribov horizon is never crossed. Thus, in presence of the condensate, the Faddeev-Popov ghost stays positive, as required by the restriction to the Gribov region. Though,  its infrared behaviour is now deeply changed. More precisely, with the inclusion of  dimension two condensate it is no more enhanced in the infrared, behaving as $\langle {\bar c} c \rangle\Big|_{k \sim 0} \sim 1/k^2$, as opposed to the enhanced behaviour $1/k^4$ observed in the absence of condensate. Needless to say, a ghost behaviour of the kind $1/k^2$ is in very good agreement with the most recent numerical lattice simulations, as well as with the studies based on the Schwinger-Dyson approach. Finally, we underline that the presence of the  dimension two condensates turns out to be energetically favoured, as shown in \cite{Dudal:2011gd} through the explicit evaluation of the effective potential.

\noindent We are now ready to evaluate the tree level diagonal gluon propagator. This will be the topic of the next section.

\section{The diagonal gluon propagator in $D=4,3,2$ dimensions} 
The four-dimensional diagonal gluon propagator can be read off from the refined action \eqref{refmag}, being given by 
\begin{equation}
\langle A_\mu(k) A_\nu(-k) \rangle_{D=4} = \left( \delta_{\mu\nu} - \frac{k_\mu k_\nu}{k^2} \right) \frac{k^2+\mu^2}{k^4 + (m^2_{\rm diag}+\mu^2)k^2 +\mu^2  \;m^2_{\rm diag}  + 4 g^2 \gamma^4}  \;, \label{diag4}
\end{equation}
where the parameters $(\mu^2,m^2_{\rm diag})$ are related to the dimension two condensates 
\begin{equation}
\mu^2 \sim \langle {\bar \phi}^{\alpha \beta}_\mu \phi^{\alpha \beta}_\mu - {\bar  \omega}^{\alpha \beta}_\mu \omega^{\alpha \beta}_\mu \rangle \;, \qquad m^2_{\rm diag} \sim \langle A^{h,3}_\mu A^{h,3}_\mu \rangle \;. \label{p}
\end{equation}
A few remarks are in order here. We notice that expression \eqref{diag4} shares a deep similarity with the gluon propagator of the Landau gauge \cite{Dudal:2007cw,Dudal:2008sp,Dudal:2011gd}. As in the Landau case, it turns out to be suppressed in the low momentum region, attaining a non-vanishing value at zero momentum, {\it i.e.} at $k^2=0$. This kind of behaviour is referred to as the decoupling solution. Moreover, expression  \eqref{diag4} is in good agreement with the lattice numerical simulations of the diagonal gluon propagator in momentum space reported in \cite{Mendes:2007vv,Mendes:2008vf}. \\\\Although not explicitly mentioned, the actions  \eqref{fmaggz1}  and \eqref{refmag} retain their validity in arbitrary dimension $D=4,3,2$, so that we can access directly the diagonal gluon propagator in both $D=3,2$. \\\\Looking at the expressions of the condensates in eqs.\eqref{cond6}, \eqref{cond7}, we see that they exist in $D=3$, the corresponding integrals being perfectly convergent in both IR and UV regions. As a consequence, for the diagonal gluon propagator in $D=3$ we get a decoupling type behaviour as well, {\it i.e.}
\begin{equation}
\langle A_\mu(k) A_\nu(-k) \rangle_{D=3} = \left( \delta_{\mu\nu} - \frac{k_\mu k_\nu}{k^2} \right) \frac{k^2+\mu^2}{k^4 + (m^2_{\rm diag}+\mu^2)k^2 + \mu^2\;m^2_{\rm diag}  + 4 g^2 \gamma^4}  \;. \label{diag3}
\end{equation}
Furthermore, from expressions \eqref{cond7}, \eqref{cond8}, one realizes that the dimension two condensates   $\langle   A^{h,3}_\mu A^{h,3}_\mu \rangle$,  $\langle {\bar \phi}^{\alpha \beta}_\mu \phi^{\alpha \beta}_\mu - {\bar  \omega}^{\alpha \beta}_\mu \omega^{\alpha \beta}_\mu \rangle$ cannot be safely introduced in $D=2$ dimensions, due to the presence of infrared singularities,  as it is apparent from the presence of the term $1/k^2$ in the integrand of expressions \eqref{cond7},\eqref{cond8}. Instead, in $D=2$ a scaling  behaviour given by a Gribov type propagator is exhibited by the diagonal gluons, namely 
\begin{equation}
\langle A_\mu(k) A_\nu(-k) \rangle_{D=2} = \left( \delta_{\mu\nu} - \frac{k_\mu k_\nu}{k^2} \right) \frac{k^2}{k^4  + 4 g^2 \gamma^4}  \;, \label{diag2}
\end{equation}
which, unlike expressions \eqref{diag4}, \eqref{diag3}, vanishes at zero momentum. Again, this feature is in full agreement with what observed in Landau gauge, where a scaling type solution emerges in $D=2$ dimensions \cite{Huber:2012zj,Dudal:2008xd,Cucchieri:2012cb}.  The emergence of a scaling type solution in $D=2$ for the diagonal propagator might have also strong consequences on the validity of the Abelian dominance hypothesis in $D=2$, as recently advocated in \cite{Gongyo:2014lxa}. 

\section{ Localization of the non-local field $A^h_\mu$ and of the BRST operator 
$s_{\gamma}^2$  } 
In the previous sections use has been made of non-local expressions, {\it i.e.} both gauge field $A^h_\mu$, eq.\eqref{hhh3g}, and BRST operator $s_{\gamma^2}$, eq.\eqref{npbrstmag}, are written in terms of non-local quantities. It is thus relevant to show here how they can be cast in a local form, so that  the standard tools of the algebraic renormalization \cite{Piguet:1995er} can be employed to analyse the structure of the theory and of its symmetry content. \\\\Let us first focus on the non-local field $A^h_\mu$ and on the non-local part of the action \eqref{fmaggz1}, which we rewrite as 
\begin{eqnarray}
S^{GZ}_{MAG} & = &  S_{YM} + S^{\gamma}_{MAG} \nonumber \\
&+& \int d^4x\left( b^{h, \alpha} D_{\mu}^{\alpha \beta}A_{\mu}^{\beta} -\bar{c}^{\alpha} \mathcal{M}^{\alpha \beta}c^{\beta} 
+g\varepsilon^{\alpha \beta}\bar{c}^{\alpha} (D_{\mu}^{\beta \delta}A_{\mu}^{\delta}) c +b^h\partial_{\mu}A_{\mu}
+\bar{c}\,\partial_{\mu}\left(\partial_{\mu}c+g\varepsilon^{\alpha \beta}A_{\mu}^{\alpha} c^{\beta} \right)\right)  \;. \nonumber \\
\label{maga1} 
\end{eqnarray}
where $S^{\gamma}_{MAG}$ stands for the non-local expression 
\begin{equation}
S^{\gamma}_{MAG}=  \int d^4x\left(\bar{\phi}^{\alpha \beta}_{\mu}\mathcal{M}^{\alpha \delta}(A^h) \phi^{\delta \beta}_{\mu}
-\bar{\omega}^{\alpha \beta}_\mu \mathcal{M}^{\alpha \delta}(A^h) \omega^{\delta \beta }_\mu + g\gamma^2 \varepsilon^{\alpha \beta}\left(\phi^{\alpha \beta }_\mu -\bar{\phi}^{\alpha \beta}_\mu \right)A^{h,3}_{\mu}  \right)  \;.
\label{fmaga2} 
\end{equation} 
The field $A^h_\mu$ can be localized by introducing an auxiliary Stueckelberg field $\xi^a$ \cite{Dragon:1996tk}, namely 
\begin{eqnarray}
{\cal C}_\mu &= & {\cal C}^a_\mu T^a = h^{\dagger} \left( A^a_\mu T^a\right)  h + \frac{i}{g} h^{\dagger} \partial_\mu h    \;, \nonumber \\
h & = & e^{ig \xi^a T^a }   \;, \label{stueck}
\end{eqnarray} 
where $\xi^a$ is a local dimensionless Stueckelberg field. The local field ${\cal C}_\mu $  can be now expanded in terms of the Stueckelberg field $\xi^a$, yielding
\begin{equation}
{\cal C}^{a}_{\mu}=A^{a}_{\mu}-D^{ab}_{\mu}\xi^{b}-\frac{g}{2}f^{abc}\xi^{b}D^{cd}_{\mu}\xi^{d}+\mathcal{O}(\xi^{3})\,.
\end{equation} 
Following \cite{Dragon:1996tk}, for the BRST transformation of $h$, one gets 
\begin{equation}
s h^{ij} = -ig c^a (T^a)^{ik} h^{kj}  \;, \qquad s^2 h  = 0  \;  \label{brstst}
\end{equation}
from which the BRST transformation of the Stueckelberg field $\xi^a$ can be evaluated iteratively, giving
\begin{equation}
s \xi^a=  - c^a + \frac{g}{2} f^{abc}c^b \xi^c - \frac{g^2}{12} f^{amr} f^{mpq} c^p \xi^q \xi^r + O(g^3)    \;.
\end{equation}
It is instructive to check here explicitly the BRST invariance of the field ${\cal C}_\mu$. For this, it is better to employ a matrix notation, namely
\begin{eqnarray}
sA_\mu &=& -\partial_\mu c + ig [A_\mu, c] \;, \qquad s c = -ig c c     \;, \nonumber \\
s h & =& -igch \;, \qquad sh^{\dagger}  = ig h^{\dagger} c  \;, \label{mbrst}
\end{eqnarray}
with $A_\mu = A^a_\mu T^a$, $c=c^a T^a$, $\xi=\xi^a T^a$. From expression  \eqref{stueck}  we easily get
\begin{eqnarray}
s {\cal C}_\mu & = & ig h^{\dagger} c \;A_\mu h + h^{\dagger}  (-\partial_\mu c + ig [A_\mu, c]) h -ig h^{\dagger} A_\mu \;c h - h^{\dagger} c \partial_\mu h + h^{\dagger} \partial_\mu(c h) \nonumber \\
&=& igh^{\dagger} c A_\mu h -  h^{\dagger}  (\partial_\mu c ) h +ig  h^{\dagger} A_\mu \;c h - ig  h^{\dagger} c \;A_\mu h -ig h^{\dagger} A_\mu c h - h^{\dagger} c \partial_\mu h + h^{\dagger} (\partial_\mu c) h + h^{\dagger} c \partial_\mu h   \nonumber \\
&=& 0 \;, \label{sah}
\end{eqnarray}
which establishes the invariant character of ${\cal C}_\mu$. \\\\The non-local action $S^{\gamma}_{MAG}$, eq.\eqref{fmaga2}, can be now replaced by the equivalent local expression 
\begin{equation}
\left[ S^{\gamma}_{MAG} \right]_{loc} =  \int d^4x\left(\bar{\phi}^{\alpha \beta}_{\mu}\mathcal{M}^{\alpha \delta}({\cal C}) \phi^{\delta \beta}_{\mu}
-\bar{\omega}^{\alpha \beta}_\mu \mathcal{M}^{\alpha \delta}({\cal C}) \omega^{\delta \beta }_\mu + g\gamma^2 \varepsilon^{\alpha \beta}\left(\phi^{\alpha \beta }_\mu -\bar{\phi}^{\alpha \beta}_\mu \right){\cal C}^{3}_{\mu}  + \tau^{\alpha} \partial_\mu {\cal C}_\mu^{\alpha} +  \tau^{3} \partial_\mu {\cal C}_\mu^{3}  \right)  \;.
\label{fmaga2}  
\end{equation}
The fields $(\tau_\mu^{\alpha}, \tau^3)$ are Lagrange multipliers imposing that the fields $({\cal C}_\mu^{\alpha}, {\cal C}_\mu^{3}) $ are  transverse, {\it i.e.}
\begin{equation}
\partial_\mu {\cal C}_\mu^{\alpha}=0 \;, \qquad  \partial_\mu {\cal C}_\mu^{3}=0   \;. \label{transv} 
\end{equation}
As discussed at length in \cite{Capri:2015ixa}, see in particular eqs.(A13-A16) of Appendix A, the trasnversality conditions \eqref{transv} can be solved iteratively for $\xi$ as a function of the gauge field $A_\mu$. In fact, imposing $\partial _{\mu }{\cal C}_\mu^{\alpha}= \partial _{\mu }{\cal C}_\mu^{3} = 0$, yields
\begin{eqnarray}
\partial ^{2}\xi &=&\partial _{\mu }A+ig[\partial _{\mu }A_{\mu },\xi
]+ig[A_{\mu },\partial _{\mu }\xi ]+g^{2}\partial _{\mu }\xi A_{\mu }\xi
+g^{2}\xi \partial _{\mu }A_{\mu }\xi +g^{2}\xi A_{\mu }\partial _{\mu
}\xi   \nonumber \\
&-&\frac{g^{2}}{2}\partial _{\mu }A_{\mu }\xi ^{2}-\frac{g^{2}}{2}A_{\mu
}\partial _{\mu }\xi \xi -\frac{g^{2}}{2}A_{\mu }\xi \partial _{\mu }\xi
-\frac{g^{2}}{2}\partial _{\mu }\xi \xi A_{\mu }-\frac{g^{2}}{2}\xi
\partial _{\mu }\xi A_{\mu }-\frac{g^{2}}{2}\xi ^{2}\partial _{\mu }A_{\mu
}  \nonumber \\
&+&i\frac{g}{2}[\xi ,\partial ^{2}\xi ]+O(\xi ^{3})\;.  \label{hh2}
\end{eqnarray}
Solving iteratively, we arrive at
\begin{equation}
\xi =\frac{1}{\partial ^{2}}\partial _{\mu }A_{\mu }+i\frac{g}{\partial ^{2}%
}\left[ \partial A,\frac{\partial A}{\partial ^{2}}\right] +i\frac{g}{%
\partial ^{2}}\left[ A_{\mu },\partial _{\mu }\frac{\partial A}{\partial ^{2}%
}\right] +\frac{i}{2}\frac{g}{\partial ^{2}}\left[ \frac{\partial A}{%
\partial ^{2}},\partial A\right] +O(A^{3})\;,  \label{phi0}
\end{equation}
and thus
\begin{equation}
{\cal C}_\mu \rightarrow A^h_\mu \;, \label{cah}
\end{equation}
where 
\begin{eqnarray}
A_{\mu }^{h} &=&A_{\mu }-\frac{1}{\partial ^{2}}\partial _{\mu }\partial A-ig%
\frac{\partial _{\mu }}{\partial ^{2}}\left[ A_{\nu },\partial _{\nu }\frac{%
\partial A}{\partial ^{2}}\right] -i\frac{g}{2}\frac{\partial _{\mu }}{%
\partial ^{2}}\left[ \partial A,\frac{1}{\partial ^{2}}\partial A\right] \nonumber \\
& + & ig\left[ A_{\mu },\frac{1}{\partial ^{2}}\partial A\right] +i\frac{g}{2}%
\left[ \frac{1}{\partial ^{2}}\partial A,\frac{\partial _{\mu }}{\partial
^{2}}\partial A\right] +O(A^{3})\;,   \label{minn2}
\end{eqnarray}
so that the starting non-local expression \eqref{hhh3g} is recovered. \\\\Let us proceed now with the localization of the BRST operator $s_{\gamma}^2$, eq.\eqref{npbrstmag}.  Here, we follow the procedure already employed in \cite{Dudal:2010hj} and we introduce the following auxiliary action 
\begin{equation}
S_{aux} = \int d^4x\left( \beta^{\alpha \beta}_{\mu}\mathcal{M}^{\alpha \delta}({\cal C}) {\bar \beta}^{\delta \beta}_{\mu}
-\bar{\psi}^{\alpha \beta}_\mu \mathcal{M}^{\alpha \delta}({\cal C}) \psi^{\delta \beta }_\mu - g \varepsilon^{\alpha \beta}  {\bar \beta}^{\alpha \beta }_\mu  {\cal C}^3_\mu   \right)  \;,  
\label{auxloc}
\end{equation}
where $\left( \beta^{\alpha \beta}_{\mu}, {\bar \beta}^{\alpha \beta}_{\mu} \right)$ are a pair of bosonic fields, while $\left( \psi^{\alpha \beta}_{\mu}, {\bar \psi}^{\alpha \beta}_{\mu} \right)$ are anti-commuting. \\\\It can be easily checked that the combination $\left( \left[ S^{\gamma}_{MAG} \right]_{loc} + S_{aux} \right)$ is left invariant by the following local, nilpotent, BRST transformations $s_{loc}$: 
\begin{eqnarray}
s_{loc} A^{\alpha}_{\mu}& = & -(D^{\alpha \beta }_{\mu}c^{\beta}+g\varepsilon^{\alpha \beta}A_{\mu}^{\beta} c)\,,\qquad s_{loc} A_{\mu}=-(\partial_{\mu}c+g\varepsilon^{\alpha \beta}A_{\mu}^{\alpha} c^{\beta})\,,\nonumber\\ 
s_{loc} c^{\alpha}& = & g\varepsilon^{\alpha \beta }c^{\beta} c\,,\qquad s_{loc} c=\frac{g}{2}\varepsilon^{\alpha \beta }c^{\alpha}c^{\beta} \,, \nonumber \\
s_{loc} \bar{c}^{\alpha}& = & b^{h, \alpha}\,,\qquad s_{loc} \bar{c}=b^h \,, \nonumber \\
s_{loc} b^{h,\alpha}& = & s_{loc} b=0 \;, \nonumber \\
s_{loc} h^{ij} & = &  -ig c^a (T^a)^{ik} h^{kj}   \;, \qquad s_{loc}{\cal C}^{\alpha}_\mu = 0 \;, \qquad s_{loc} {\cal C}^3_\mu = 0 \;, \nonumber \\ 
s_{loc} \phi^{\alpha \beta}_{\mu}  &=&  \omega^{\alpha \beta}_{\mu} \;, \qquad s_{loc} \omega^{\alpha \beta}_{\mu} = 0 \;, \nonumber \\
s_{loc} {\bar \omega}^{\alpha \beta}_\mu &=& {\bar \phi}^{\alpha \beta}_\mu + \gamma^2 \beta^{\alpha \beta}_\mu \;, \qquad s_{loc} {\bar \phi}^{\alpha \beta}_\mu =0 \;, \qquad s_{loc}  \beta^{\alpha \beta}_\mu = 0 \;, \nonumber \\
s_{loc} {\bar \beta}^{\alpha \beta}_{\mu} & = & \gamma^2 {\omega}^{\alpha \beta}_{\mu}  \;, \qquad s_{loc} {\bar \psi}^{\alpha \beta}_{\mu} =0 \;, \qquad s_{loc} { \psi}^{\alpha \beta}_{\mu} =0 \;, 
\label{brstloc}
\end{eqnarray}
and 
\begin{equation}
s_{loc} s_{loc} =0 \;, \qquad s_{loc} \left( \left[ S^{\gamma}_{MAG} \right]_{loc} + S_{aux} \right) =0 \;.    \label{exloc}
\end{equation}
In fact, 
\begin{eqnarray}
s_{loc} \left( \left[ S^{\gamma}_{MAG} \right]_{loc} + S_{aux} \right)  & = &  \int d^4x \left( {\bar \phi}^{\alpha \beta}_{\mu}\mathcal{M}^{\alpha \delta}({\cal C}) {\omega}^{\delta \beta}_{\mu}  
- ( \bar{\phi}^{\alpha \beta}_\mu +\gamma^2 \beta^{\alpha \beta}_\mu) \mathcal{M}^{\alpha \delta}({\cal C}) \omega ^{\delta \beta }_\mu +  g \gamma^2 \varepsilon^{\alpha \beta}  { \omega}^{\alpha \beta }_\mu  {\cal C}^3_\mu   \right)   \nonumber \\
&+& \int d^4x \left( \gamma^2   \beta^{\alpha \beta}_\mu \mathcal{M}^{\alpha \delta}({\cal C})  {\omega}^{\delta \beta}_{\mu}   -    g \gamma^2 \varepsilon^{\alpha \beta}  { \omega}^{\alpha \beta }_\mu  {\cal C}^3_\mu \right)  \nonumber \\
&=& 0 \;.   \label{invloc}
\end{eqnarray}
As explained in \cite{Dudal:2010hj}, the introduction  of the auxiliary action \eqref{auxloc} enables us to localize the BRST operator $s_{\gamma^2}$. This follows by looking at the equation of motion of the auxiliary field ${\bar \beta}$, namely
\begin{equation}
\frac{\delta S_{aux}}{\delta \bar \beta} = \mathcal{M}({\cal C}) \beta - g  {\cal C}^3 =0 \qquad \Rightarrow   \qquad \beta = g \;{\cal C}^3 \frac{1}{ \mathcal{M}({\cal C})}  \;. 
\end{equation}
Therefore, from equation \eqref{cah}, one gets 
\begin{eqnarray}
s_{loc} {\bar \omega} & = & {\bar \phi} + \gamma^2 \beta \qquad \Rightarrow  \qquad s_{\gamma^2} {\bar \omega}  = {\bar \phi} +   g \gamma^2 \;{ (A^h)}^3 \frac{1}{ \mathcal{M}({A^h})}  \;, \nonumber \\
s_{loc} & \Rightarrow  & s_{\gamma^2} \;,  \label{slsg}
\end{eqnarray}
so that the non-local expression of the operator $s_{\gamma}$, eq.\eqref{npbrstmag}, is recovered. \\\\In summary, the localized action\footnote{We also point out that the Lagrange multipliers  $(b^h, b^{h,\alpha})$ appearing in expression \eqref{maga3} can be considered as elementary fields. This follows by noticing that the field redefinitions in eq.\eqref{redb} correspond in fact to field transformations with unit Jacobian.} 
\begin{eqnarray}
\left[ S^{GZ}_{MAG} \right]_{loc}& = &  S_{YM} + \left[ S^{\gamma}_{MAG} \right]_{loc} + S_{aux}  \nonumber \\
&+& \int d^4x\left( b^{h, \alpha} D_{\mu}^{\alpha \beta}A_{\mu}^{\beta} -\bar{c}^{\alpha} \mathcal{M}^{\alpha \beta}c^{\beta} 
+g\varepsilon^{\alpha \beta}\bar{c}^{\alpha} (D_{\mu}^{\beta \delta}A_{\mu}^{\delta}) c +b^h\partial_{\mu}A_{\mu}
+\bar{c}\,\partial_{\mu}\left(\partial_{\mu}c+g\varepsilon^{\alpha \beta}A_{\mu}^{\alpha} c^{\beta} \right)\right)  \;, \nonumber \\
\label{maga3} 
\end{eqnarray}
is left invariant by the local, nilpotent, BRST transformations \eqref{brstloc}, {\it i.e.}
\begin{equation}
s_{loc} \left[ S^{GZ}_{MAG} \right]_{loc} = 0 \;, \qquad s_{loc} s_{loc} = 0 \;.    \label{loc1}
\end{equation}
Both action \eqref{maga3}  and BRST operator \eqref{brstloc}  reduce to their non-local expressions when the auxiliary localizing fields $(\xi, {\bar \beta}, \beta, {\psi}, {\bar \psi})$ are integrated out. Moreover, equations \eqref{loc1} can be immediately translated into Slavnov-Taylor identities useful for studying the algebraic renormalization as well as the cohomology of the local BRST operator $s_{loc}$. We shall post-pone this study to a further detailed analysis. Let us end this section by calling attention to the fact that the non-polynomial character of the local action $ \left[ S^{GZ}_{MAG} \right]_{loc}$ in the Stueckelberg field $\xi$ does not jeopardise the use of the powerful tools of the algebraic renormalization  \cite{Piguet:1995er}, which have been already successfully applied to other non-polynomial actions like, for instance, $N=1$ Super-Yang-Mills in super-space, chiral Wess-Zumino models in two dimensions as well as non-linear sigma models.

\section{Conclusion} 
In this work the generalization of the non-perturbative exact nilpotent symmetry of the Gribov-Zwanziger action  constructed in \cite{Capri:2015ixa} in the Landau gauge has been generalized to the case of Yang-Mills theories quantized in the maximal Abelian gauge, as summarized by eqs.\eqref{npbrstmag}, \eqref{exmag}. \\\\It is worth to point out the deep similarity existing between the non-perturbative nilpotent BRST operators of the Landau and maximal Abelian gauges, as one can infer by comparing  expressions \eqref{brst3} and \eqref{npbrstmag}. One notices  in fact that the Faddeev-Popov operators of both gauges appear in exactly the same way, {\it i.e} in the transformation of the auxiliary field $\omega$. The presence of the inverse of the Faddeev-Popov operators shows that the non-perturbative BRST operator $s_{\gamma^2}$ feels the presence  of the Gribov horizon. As such, the operator $s_{\gamma^2}$ is deeply intertwined with the geometry of the corresponding Gribov regions in both gauges.  \\\\The exact non-perturbative BRST symmetry has also been extended to the refined version of the theory, eq.\eqref{refmag}, which takes into account the existence of the dimension two condensates.\\\\The resulting diagonal gluon propagator has been evaluated in $D=4,3,2$ dimensions. While in $D=4,3$ a decoupling type behaviour has been found, see eqs.\eqref{diag3}, \eqref{diag3}, in $D=2$ a scaling type behaviour  emerges, as given in eq.\eqref{diag2}. Once more, we underline the strict analogy existing with the gluon propagator of the Landau gauge, which exhibits a similar  behavior. This feature suggests a kind of general behavior  of the gluon propagator in different gauges, as recently advocated in \cite{Guimaraes:2015bra}, where a study of the equal-time spatial gluon propagator has been performed in the Coulomb gauge, obtaining similar results. \\\\Finally, we hope that this work will stimulate our colleagues from the lattice community to pursue the numerical studies of  the diagonal gluon propagator in both $D=4,3,2$ dimensions.  

\section*{Acknowledgements}

The Conselho Nacional de Desenvolvimento Cient\'{\i}fico e
Tecnol\'{o}gico (CNPq-Brazil), the FAPERJ, Funda{\c{c}}{\~{a}}o de
Amparo {\`{a}} Pesquisa do Estado do Rio de Janeiro,  the
Coordena{\c{c}}{\~{a}}o de Aperfei{\c{c}}oamento de Pessoal de
N{\'{\i}}vel Superior (CAPES),  are gratefully acknowledged.

\end{document}